# AN ALGORITHM FOR FAST COMPUTATION OF THE MULTIRESOLUTION DISCRETE FOURIER TRANSFORM

Bartosz ANDREATTO, Aleksandr CARIOW

West Pomeranian University of Technology, Szczecin, Department of Computer Architectures and Telecommunications
Corresponding author: Bartosz ANDREATTO, E-mail: `bandreatto@gmail.com`

The article presents a computationally effective algorithm for calculating the multiresolution discrete Fourier transform (MrDFT). The algorithm is based on the idea of reducing the computational complexity which was introduced by Wen and Sandler [10] and utilizes the vectorization of calculating process at each stage of the considered transformation. This allows for the use of a computational process parallelization and results in a reduction of computation time. In the description of the computational procedure, which describes the algorithm, we use the matrix notation. This notation enables to represent adequately the space-time structures of the implemented computational process and directly map these structures into the constructions of a high-level programming language or into a hardware realization space.

*Key words*: multiresolution discrete Fourier transform, fast Fourier transform, multiresolution analysis, multiresolution domain transformations.

## 1. INTRODUCTION

In recent years we could experience a lot of publications regarding applications of the multiresolution discrete Fourier transform [2, 3, 5, 6, 7]. The issues of computational rationalization of this transform were also discussed. The paper [1] demonstrates the possibility of the reduction of the computational complexity in calculating the MrDFT through the use of the FFT algorithm in the calculation of the spectral components for each resolution level. However, in the presented approach not all of the possibilities for the reduction of the excessive calculations were taken into account. The article [10] describes another efficient way of computing the multiresolution discrete Fourier transform. This approach also provides for the use of the FFT algorithm at each stage of calculating the spectral MrDFT components. Nonetheless, in contrast to previous approach there is a possibility for calculating the spectral components of the current level of the resolution on the basis of the results obtained during the previous calculations. By using presented in this study ways of reducing the computational complexity, we developed a rationalized algorithm to compute the MrDFT which is based on the vectorization of a computational process at each stage of its realization. The computational procedure describing this algorithm is presented in a form of vector-matrix transformations. The formulation of this procedure by means of matrix algebra constructions offers not only a comfortable formalism for describing the algorithm but also enables the implementation of computations by simple vector-matrix manipulations that are well-suited to be implemented in VLSI signal processor units with a scalable level of parallelism. Knowledge of matrix components structures and their positions in the computational procedure allows us to define the composition and functionality of separate processor modules as well as to perform useful prerequisites for the effective implementation of the MrDFT processors in the common VLSI circuit. Moreover, these algebraic constructions can be directly used for fast implementation in the high-level matrix-oriented languages like Matlab.

## 2. THE DEFINITION OF THE VECTORIZED FORM OF THE MULTIRESOLUTION DISCRETE FOURIER TRANSFORM

Let us consider the one-dimensional discrete signal whose length can be expressed as $N = 2^m$, for $m \in \mathbb{N}$. This signal can be represented by the following column vector:

$$\mathbf{X}_{N \times 1} = [x_0, x_1, ..., x_{N-1}]^T.$$

The vectorized form of the multiresolution discrete Fourier transform is defined as follows [1]:

$$\mathbf{Y}_{mN \times 1} = [\bigoplus_{i=1}^{m} (\mathbf{I}_{2^{m-1}} \otimes \mathbf{E}_{2^i})] \mathbf{P}_{mN \times N} \mathbf{X}_{N \times 1}, \quad (1)$$

where $\mathbf{Y}_{mN \times 1} = [\mathbf{Y}^{(1)}_{N \times 1}, \mathbf{Y}^{(2)}_{N \times 1}, ..., \mathbf{Y}^{(m)}_{N \times 1}]^T$ - is an output column vector whose elements are vectors, $\mathbf{Y}^{(i)}_{N \times 1} = [y^{(i)}_0, y^{(i)}_1, ..., y^{(i)}_{N-1}]^T$ - is a column vector defining the spectral components at the $i$-th resolution level, and $\mathbf{E}_{2^i}$ - is a matrix of values of the discrete exponential functions (DEF) [4] with the elements defined as $w^{n,k}_{2^i} = \exp(-\mathrm{j} \cdot 2\pi nk / 2^i)$ where $\mathrm{j}$ is an imaginary unit satisfying $\mathrm{j}^2 = -1$. This matrix takes the following form:

$$\mathbf{E}_{2^i} = \begin{bmatrix} w^{0,0}_{2^i} & w^{0,1}_{2^i} & \cdots & w^{0,2^i-1}_{2^i} \\ w^{1,0}_{2^i} & w^{1,1}_{2^i} & \cdots & w^{1,2^i-1}_{2^i} \\ \vdots & \vdots & \ddots & \vdots \\ w^{2^i-1,0}_{2^i} & w^{2^i-1,1}_{2^i} & \cdots & w^{2^i-1,2^i-1}_{2^i} \end{bmatrix}.$$

$\mathbf{I}_N$ is an identity $N \times N$ matrix, in turn the symbols $\otimes$ and $\oplus$ indicate a tensor product and a direct sum of two matrices respectively [9]. The matrix $\mathbf{P}_{mN \times N}$ is defined as:

$$\mathbf{P}_{mN \times N} = \mathbf{1}_{m \times 1} \otimes \mathbf{I}_N,$$

where $\mathbf{1}_{m \times 1}$ is the $m \times 1$ matrix whose elements are units.

It can be seen from the formula (1), that at the $i$-th ($i = 1, 2, ..., m$) level of the multiresolution discrete Fourier transform there are $2^{m-1}$ sub-blocks of vector-matrix multiplications denoted by $\mathbf{E}_{2^i}$. Thus, the $i$-th level of the MrDFT may be regarded as a short-time Fourier transform with the rectangular window whose length is strictly defined for a given level and equals $2^i$.

## 3. DEVELOPMENT OF THE RATIONALIZED ALGORITHM FOR CALCULATING THE MULTIRESOLUTION DISCRETE FOURIER TRANSFORM

The approach presented in this article consists in the realization of the fast Fourier transform without any decimation and making use of the internal results during the calculation of consecutive levels forming the multiresolution discrete Fourier transform. In the rest of this section it will be presented a synthesis of the rationalized algorithm for a fast computation of the MrDFT. Let us introduce some auxiliary matrix structures that will be used in the synthesis of the final computational procedure.

First we introduce the "matrix of data expanding" for the first iteration of the algorithm:

$$\mathbf{P}^{(1)}_{mN \times N} = \mathbf{1}_{m \times 1} \otimes \mathbf{I}_N,$$

For the $i$-th iteration ($i = 2, 3, ..., m$) it may be written as:

$$\mathbf{P}^{(i)}_{(m+1)N \times mN} = \mathbf{I}_{(i-2)N} \oplus (\mathbf{1}_{2 \times 1} \otimes \mathbf{I}_N) \oplus \mathbf{I}_{(m-i+1)N}.$$

Next we introduce the "algebraic summation matrix" for $i = 1$ that is defined in the following way:

$$\mathbf{A}_{mN}^{(1)} = (\mathbf{I}_{2^{m-1}} \otimes \mathbf{H}_2) \oplus \mathbf{I}_{(m-1)N},$$

while for $(i = 2, 3, ..., m)$ it can be written as:

$$\mathbf{A}_{mN \times (m+1)N}^{(i)} = \mathbf{I}_{(i-1)N} \oplus [\mathbf{U}_N^{(i)} \mid \mathbf{V}_N^{(i)}] \oplus \mathbf{I}_{(m-i)N},$$

where $\mathbf{H}_2$ is the $2 \times 2$ Hadamard matrix. The matrices $\mathbf{U}_N^{(i)}$ and $\mathbf{V}_N^{(i)}$ are the matrices of algebraic additions and subtractions taking the following forms:

$$\mathbf{U}_N^{(i)} = \mathbf{I}_{2^{m-i}} \otimes \left[ \begin{array}{c} \mathbf{1}_{1\times 2} \otimes \mathbf{I}_{2^{i-1}} \\ \hline \mathbf{0}_{2^{i-1} \times 2^i} \end{array} \right], \qquad (2)$$

$$\mathbf{V}_N^{(i)} = \mathbf{I}_{2^{m-i}} \otimes \left[ \begin{array}{c} \mathbf{0}_{2^{i-1} \times 2^i} \\ \hline \mathbf{I}_{2^{i-1}} \mid -\mathbf{I}_{2^{i-1}} \end{array} \right], \qquad (3)$$

where $\mathbf{0}_{N \times K}$ is a zero matrix whose size is $N \times K$.

The figure 1 presents the data-flow diagrams showing ways of multiplying the matrices $\mathbf{U}_N^{(i)}$ and $\mathbf{V}_N^{(i)}$ by a column vector of temporary data in accordance with the formulas (2) and (3). In this figure, the following designations were adopted:

$$\widetilde{\mathbf{V}}_{2^i}^{(i)} = \left[ \begin{array}{c} \mathbf{0}_{2^{i-1} \times 2^i} \\ \hline \mathbf{I}_{2^{i-1}} \mid -\mathbf{I}_{2^{i-1}} \end{array} \right], \; \widetilde{\mathbf{V}}_{2^i}^{(i)} = \left[ \begin{array}{c} \mathbf{0}_{2^{i-1} \times 2^i} \\ \hline \mathbf{I}_{2^{i-1}} \mid -\mathbf{I}_{2^{i-1}} \end{array} \right].$$

In this article all the data-flow diagrams are oriented from left to right. Straight lines in the figures denote the operation of data transfer. The rectangles indicate the operation of multiplication by the matrix inscribed inside an element. In the figures presented in this paper we denote the $j$-th element of an input vector by $\tilde{x}_j$, and by $\tilde{y}_j$ we specify the $j$-th element of an output vector. At points where lines converge, the data are summarized. The dashed lines indicate the operation of subtraction.

Then we introduce the matrices specifying the positions and sizes of "butterflies" for $(i = 1, 2, ..., m)$ that are defined as follows:

$$\mathbf{D}_{mN}^{(i)} = \mathbf{I}_{(i-1)N} \oplus (\mathbf{I}_{2^{m-i}} \otimes \widetilde{\mathbf{D}}_{2^i}^{(i)}) \oplus \mathbf{I}_{(m-i)N},$$

where

$$\widetilde{\mathbf{D}}_{2^i}^{(i)} = \mathbf{I}_{2^{i-1}} \oplus \mathrm{diag}(w_{2^i}^0, w_{2^i}^1, ..., w_{2^i}^{2^{i-1}-1}).$$

In the above expression, the quantity $w_n^k = \exp(-\mathrm{j} \cdot 2\pi k / n)$ is called a twiddle factor [8].

Subsequently, we introduced matrix $\mathbf{F}_{mN}^{(1)}$, that is defined in the following way:

$$\mathbf{F}_{mN}^{(1)} = \mathbf{I}_{mN},$$

in turn for $(i = 2, 3, ..., m)$ we can write:

$$\mathbf{F}_{mN}^{(i)} = \mathbf{I}_{(i-1)N} \oplus (\mathbf{I}_{2^{m-i}} \otimes \widetilde{\mathbf{R}}_{2^i}^{(i)}) \oplus \mathbf{I}_{(m-i)N}, \; \widetilde{\mathbf{R}}_{2^i}^{(i)} = \mathbf{I}_{2^{i-1}} \oplus \mathbf{R}_{2^{i-1}}^{(i)}.$$

The elements of the matrix $\mathbf{R}_{2^{i-1}}^{(i)}$ for $(i = 2, 3, ..., m)$ can be computed as follows:

$$\mathbf{R}_{2^{i-1}}^{(i)} = (\mathbf{W}_{2^{i-1}}^{i-1} \mathbf{S}_{2^{i-1}}^{i-1})(\mathbf{W}_{2^{i-1}}^{i-2} \mathbf{S}_{2^{i-1}}^{i-2}) \times \cdots \times (\mathbf{W}_{2^{i-1}}^{1} \mathbf{S}_{2^{i-1}}^{1}),$$

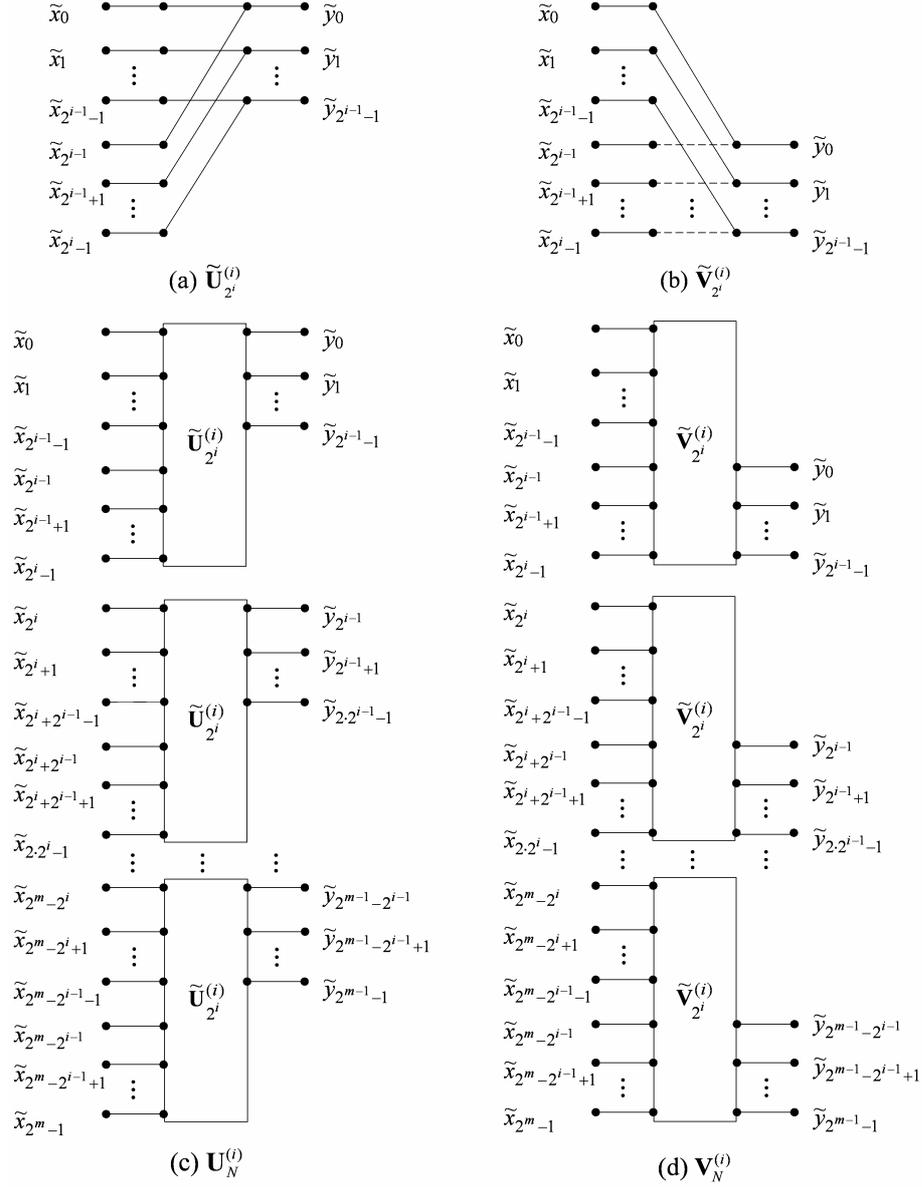

Fig. 1 – The data-flow diagrams showing ways of multiplying the matrices $\mathbf{U}_N^{(i)}$ and $\mathbf{V}_N^{(i)}$ by a column vector of temporary data in accordance with the formulas (2) and (3).

where the matrices $\mathbf{S}_{2^{i-1}}^{(j)}$ and $\mathbf{W}_{2^{i-1}}^{(j)}$ for ($j = 1, 2, ..., i-1$) take the following forms:

$$\mathbf{S}_{2^{i-1}}^{(j)} = \mathbf{I}_{2^{j-1}} \otimes (\mathbf{H}_2 \otimes \mathbf{I}_{2^{i-j-1}}), \quad \mathbf{W}_{2^{i-1}}^{(j)} = \mathbf{I}_{2^{j-1}} \otimes [\mathbf{I}_{2^{i-j-1}} \oplus \operatorname{diag}(w_{2^{i-j}}^0, w_{2^{i-j}}^1, ..., w_{2^{i-j}}^{2^{i-j-1}-1})].$$

At last we introduce the permutation matrix $\mathbf{\Gamma}_{mN}$, that is defined as:

$$\mathbf{\Gamma}_{mN} = \bigoplus_{i=1}^{m}(\mathbf{I}_{2^{m-i}} \otimes \mathbf{T}_{2^i}^{(i)}),$$

where by $\mathbf{T}_{2^i}^{(i)} = [t_{k,n}]$ we denote the orthogonal matrix whose nonzero elements are defined as:

$$t_{k,\langle n-1\rangle_2+1} = 1, \text{ for } k = n = \overline{1, 2^i},$$

where $\langle p \rangle_2$ means a binary inversion of a number $p = \sum_{i=0}^{m-1}(p_i \cdot 2^i)$ [8].

Taking into account all the above assumptions and methods of the construction of the particular matrix structures, we can write a common procedure for the rationalized computation of the multiresolution discrete Fourier transform:

$$\mathbf{Y}_{mN \times 1} = \mathbf{\Gamma}_{mN}(\mathbf{F}_{mN}^{(m)}\mathbf{D}_{mN}^{(m)}\mathbf{A}_{mn \times (m+1)N}^{(m)}\mathbf{P}_{(m+1)N \times mN}^{(m)})(\mathbf{F}_{mN}^{(m-1)}\mathbf{D}_{mN}^{(m-1)}\mathbf{A}_{mn \times (m+1)N}^{(m-1)}\mathbf{P}_{(m+1)N \times mN}^{(m-1)}) \times \cdots \\ \times (\mathbf{F}_{mN}^{(1)}\mathbf{D}_{mN}^{(1)}\mathbf{A}_{mn}^{(1)}\mathbf{P}_{mN \times N}^{(1)})\mathbf{X}_{N \times 1}. \tag{4}$$

## 4. AN EXAMPLE OF CALCULATIONS

Let us consider the synthesis of the rationalized algorithm for computing the MrDFT for $N = 8$ ($m = 3$) in accordance with the formula (4). The computational procedure for this example takes the following form:

$$\mathbf{Y}_{24 \times 1} = \mathbf{\Gamma}_{24}(\mathbf{F}_{24}^{(3)}\mathbf{D}_{24}^{(3)}\mathbf{A}_{24 \times 32}^{(3)}\mathbf{P}_{32 \times 24}^{(3)})(\mathbf{F}_{24}^{(2)}\mathbf{D}_{24}^{(2)}\mathbf{A}_{24 \times 32}^{(2)}\mathbf{P}_{32 \times 24}^{(2)})(\mathbf{F}_{24}^{(1)}\mathbf{D}_{24}^{(1)}\mathbf{A}_{24}^{(1)}\mathbf{P}_{24 \times 8}^{(1)})\mathbf{X}_{8 \times 1}, \tag{5}$$

where the input signal is represented as follows:

$$\mathbf{X}_{8 \times 1} = [x_0, x_1, x_2, x_3, x_4, x_5, x_6, x_7]^T,$$

in turn the output vector may be defined as:

$$\mathbf{Y} = [\mathbf{Y}_{8 \times 1}^{(1)}, \mathbf{Y}_{8 \times 1}^{(2)}, \mathbf{Y}_{8 \times 1}^{(3)}]^T,$$

where for $i = 1, 2, 3$ we can write:

$$\mathbf{Y}_{8 \times 1}^{(i)} = [y_0^{(i)}, y_1^{(i)}, \ldots, y_7^{(i)}]^T.$$

In this case, the respective matrix constructions are formed as follows:

$$\mathbf{P}_{24 \times 8}^{(1)} = \mathbf{1}_{3 \times 1} \otimes \mathbf{I}_8, \; \mathbf{A}_{24}^{(1)} = (\mathbf{I}_4 \otimes \mathbf{H}_2) \oplus \mathbf{I}_{16}, \; \mathbf{D}_{24}^{(1)} = (\mathbf{I}_4 \otimes \tilde{\mathbf{D}}_2^{(1)}) \oplus \mathbf{I}_{16}, \; \tilde{\mathbf{D}}_2^{(1)} = \mathbf{I}_1 \oplus w_2^0, \; \mathbf{F}_{24}^{(1)} = \mathbf{I}_{24},$$

$$\mathbf{P}_{32 \times 24}^{(2)} = (\mathbf{1}_{2 \times 1} \otimes \mathbf{I}_8) \oplus \mathbf{I}_{16} = \begin{bmatrix} \mathbf{I}_8 & \mathbf{0}_8 & \mathbf{0}_8 \\ \mathbf{I}_8 & \mathbf{0}_8 & \mathbf{0}_8 \\ \mathbf{0}_8 & & \\ \mathbf{0}_8 & \mathbf{I}_{16} & \end{bmatrix}, \; \mathbf{A}_{24 \times 32}^{(2)} = \mathbf{I}_8 \oplus [\mathbf{U}_8^{(2)} | \mathbf{V}_8^{(2)}] \oplus \mathbf{I}_8 = \begin{bmatrix} \mathbf{I}_8 & \mathbf{0}_{8 \times 16} & \mathbf{0}_8 \\ \mathbf{0}_8 & \mathbf{U}_8^{(2)} | \mathbf{V}_8^{(2)} & \mathbf{0}_8 \\ \mathbf{0}_8 & \mathbf{0}_{8 \times 16} & \mathbf{I}_8 \end{bmatrix},$$

$$\mathbf{U}_8^{(2)} = \mathbf{I}_2 \otimes \begin{bmatrix} \mathbf{1}_{1 \times 2} \otimes \mathbf{I}_2 \\ \mathbf{0}_{2 \times 4} \end{bmatrix} = \begin{bmatrix} \mathbf{I}_2 | \mathbf{I}_2 & \mathbf{0}_4 \\ \mathbf{0}_{2 \times 4} & \\ \mathbf{0}_4 & \mathbf{I}_2 | \mathbf{I}_2 \\ & \mathbf{0}_{2 \times 4} \end{bmatrix}, \; \mathbf{V}_8^{(2)} = \mathbf{I}_2 \otimes \begin{bmatrix} \mathbf{0}_{2 \times 4} \\ \mathbf{I}_2 | -\mathbf{I}_2 \end{bmatrix} = \begin{bmatrix} \mathbf{0}_{2 \times 4} & \mathbf{0}_4 \\ \mathbf{I}_2 | -\mathbf{I}_2 & \\ \mathbf{0}_4 & \mathbf{0}_{2 \times 4} \\ & \mathbf{I}_2 | -\mathbf{I}_2 \end{bmatrix},$$

$$\mathbf{D}_{24}^{(2)} = \mathbf{I}_8 \oplus (\mathbf{I}_2 \otimes \tilde{\mathbf{D}}_4^{(2)}) \oplus \mathbf{I}_8 = \begin{bmatrix} \mathbf{I}_8 & \mathbf{0}_8 & \begin{matrix}\mathbf{0}_4 & \mathbf{0}_4 \\ \mathbf{0}_4 & \mathbf{0}_4\end{matrix} \\ \mathbf{0}_8 & \begin{matrix}\tilde{\mathbf{D}}_4^{(2)} & \mathbf{0}_4 \\ \mathbf{0}_4 & \tilde{\mathbf{D}}_4^{(2)}\end{matrix} & \mathbf{0}_8 \\ \begin{matrix}\mathbf{0}_4 & \mathbf{0}_4 \\ \mathbf{0}_4 & \mathbf{0}_4\end{matrix} & \mathbf{0}_8 & \mathbf{I}_8 \end{bmatrix}, \; \tilde{\mathbf{D}}_4^{(2)} = \mathbf{I}_2 \oplus \mathrm{diag}(w_4^0, w_4^1),$$

$$\mathbf{F}_{24}^{(2)} = \mathbf{I}_8 \oplus (\mathbf{I}_2 \otimes \widetilde{\mathbf{R}}_4^{(2)}) \oplus \mathbf{I}_8 = \begin{bmatrix} \mathbf{I}_8 & \mathbf{0}_8 & \begin{matrix}\mathbf{0}_4 & \mathbf{0}_4 \\ \mathbf{0}_4 & \mathbf{0}_4\end{matrix} \\ \mathbf{0}_8 & \begin{matrix}\widetilde{\mathbf{R}}_4^{(2)} & \mathbf{0}_4 \\ \mathbf{0}_4 & \widetilde{\mathbf{R}}_4^{(2)}\end{matrix} & \mathbf{0}_8 \\ \begin{matrix}\mathbf{0}_4 & \mathbf{0}_4 \\ \mathbf{0}_4 & \mathbf{0}_4\end{matrix} & \mathbf{0}_8 & \mathbf{I}_8 \end{bmatrix}, \widetilde{\mathbf{R}}_4^{(2)} = \mathbf{I}_2 \oplus \mathbf{R}_2^{(2)},$$

$$\mathbf{R}_2^{(2)} = \mathbf{W}_2^{(1)}\mathbf{S}_2^{(1)}, \mathbf{W}_2^{(1)} = \mathbf{I}_1 \oplus w_2^0, \mathbf{S}_2^{(1)} = \mathbf{H}_2,$$

$$\mathbf{P}_{32\times 24}^{(3)} = \mathbf{I}_8 \oplus (\mathbf{1}_{2\times 1} \otimes \mathbf{I}_8) \oplus \mathbf{I}_8 = \begin{bmatrix} \mathbf{I}_8 & \mathbf{0}_8 & \mathbf{0}_8 \\ \mathbf{0}_8 & \mathbf{I}_8 & \mathbf{0}_8 \\ \mathbf{0}_8 & \mathbf{I}_8 & \mathbf{0}_8 \\ \mathbf{0}_8 & \mathbf{0}_8 & \mathbf{I}_8 \end{bmatrix}, \mathbf{A}_{24\times 32}^{(3)} = \mathbf{I}_{16} \oplus [\mathbf{U}_8^{(3)} | \mathbf{V}_8^{(3)}] = \begin{bmatrix} \mathbf{I}_{16} & \begin{matrix}\mathbf{0}_8 & \mathbf{0}_8 \\ \mathbf{0}_8 & \mathbf{0}_8\end{matrix} \\ \begin{matrix}\mathbf{0}_8 & \mathbf{0}_8\end{matrix} & \begin{matrix}\mathbf{U}_8^{(3)} & \mathbf{V}_8^{(3)}\end{matrix} \end{bmatrix},$$

$$\mathbf{U}_8^{(3)} = \begin{bmatrix} \mathbf{1}_{1\times 2} \otimes \mathbf{I}_4 \\ \mathbf{0}_{4\times 8} \end{bmatrix} = \begin{bmatrix} \mathbf{I}_4 & \mathbf{I}_4 \\ \mathbf{0}_{4\times 8} \end{bmatrix}, \mathbf{V}_8^{(3)} = \begin{bmatrix} \mathbf{0}_{4\times 8} \\ \mathbf{I}_4 & -\mathbf{I}_4 \end{bmatrix},$$

$$\mathbf{D}_{24}^{(3)} = \mathbf{I}_{16} \oplus \widetilde{\mathbf{D}}_8^{(3)} = \begin{bmatrix} \mathbf{I}_8 & \mathbf{0}_8 & \mathbf{0}_8 \\ \mathbf{0}_8 & \mathbf{I}_8 & \mathbf{0}_8 \\ \mathbf{0}_8 & \mathbf{0}_8 & \widetilde{\mathbf{D}}_8^{(3)} \end{bmatrix}, \widetilde{\mathbf{D}}_8^{(3)} = \mathbf{I}_4 \oplus \mathrm{diag}(w_8^0, w_8^1, w_8^2, w_8^3),$$

$$\mathbf{F}_{24}^{(3)} = \mathbf{I}_{16} \oplus \widetilde{\mathbf{R}}_8^{(3)} = \begin{bmatrix} \mathbf{I}_8 & \mathbf{0}_8 & \mathbf{0}_8 \\ \mathbf{0}_8 & \mathbf{I}_8 & \mathbf{0}_8 \\ \mathbf{0}_8 & \mathbf{0}_8 & \widetilde{\mathbf{R}}_8^{(3)} \end{bmatrix}, \widetilde{\mathbf{R}}_8^{(3)} = \mathbf{I}_4 \oplus \mathbf{R}_4^{(3)}, \mathbf{R}_4^{(3)} = (\mathbf{W}_4^{(2)}\mathbf{S}_4^{(2)})(\mathbf{W}_4^{(1)}\mathbf{S}_4^{(1)}),$$

$$\mathbf{S}_4^{(1)} = \mathbf{H}_2 \otimes \mathbf{I}_2, \mathbf{W}_4^{(1)} = \mathbf{I}_2 \oplus \mathrm{diag}(w_4^0, w_4^1), \mathbf{S}_4^{(2)} = \mathbf{I}_2 \otimes \mathbf{H}_2, \mathbf{W}_4^{(2)} = \mathbf{I}_2 \otimes (\mathbf{I}_1 \oplus w_2^0),$$

$$\mathbf{\Gamma}_{24} = \bigoplus_{i=1}^{3}(\mathbf{I}_{2^{3-i}} \otimes \mathbf{T}_{2^i}^{(i)}) = \begin{bmatrix} \mathbf{I}_4 \otimes \mathbf{T}_2^{(1)} & \mathbf{0}_8 & \mathbf{0}_8 \\ \mathbf{0}_8 & \mathbf{I}_2 \otimes \mathbf{T}_4^{(2)} & \mathbf{0}_8 \\ \mathbf{0}_8 & \mathbf{0}_8 & \mathbf{I}_1 \otimes \mathbf{T}_8^{(3)} \end{bmatrix},$$

$$\mathbf{T}_2^{(1)} = \begin{bmatrix} 1 & 0 \\ 0 & 1 \end{bmatrix}, \mathbf{T}_4^{(2)} = \begin{bmatrix} 1 & 0 & 0 & 0 \\ 0 & 0 & 1 & 0 \\ 0 & 1 & 0 & 0 \\ 0 & 0 & 0 & 1 \end{bmatrix}, \mathbf{T}_8^{(3)} = \begin{bmatrix} 1 & 0 & 0 & 0 & 0 & 0 & 0 & 0 \\ 0 & 0 & 0 & 0 & 1 & 0 & 0 & 0 \\ 0 & 0 & 1 & 0 & 0 & 0 & 0 & 0 \\ 0 & 0 & 0 & 0 & 0 & 0 & 1 & 0 \\ 0 & 1 & 0 & 0 & 0 & 0 & 0 & 0 \\ 0 & 0 & 0 & 0 & 0 & 1 & 0 & 0 \\ 0 & 0 & 0 & 1 & 0 & 0 & 0 & 0 \\ 0 & 0 & 0 & 0 & 0 & 0 & 0 & 1 \end{bmatrix}.$$

The figure 2 shows the data-flow diagram of the MrDFT calculation process for $N = 8$ in accordance with the procedure given by the formula (5). The circles in this figure denote the operation of multiplication by a twiddle factor inscribed inside an element.

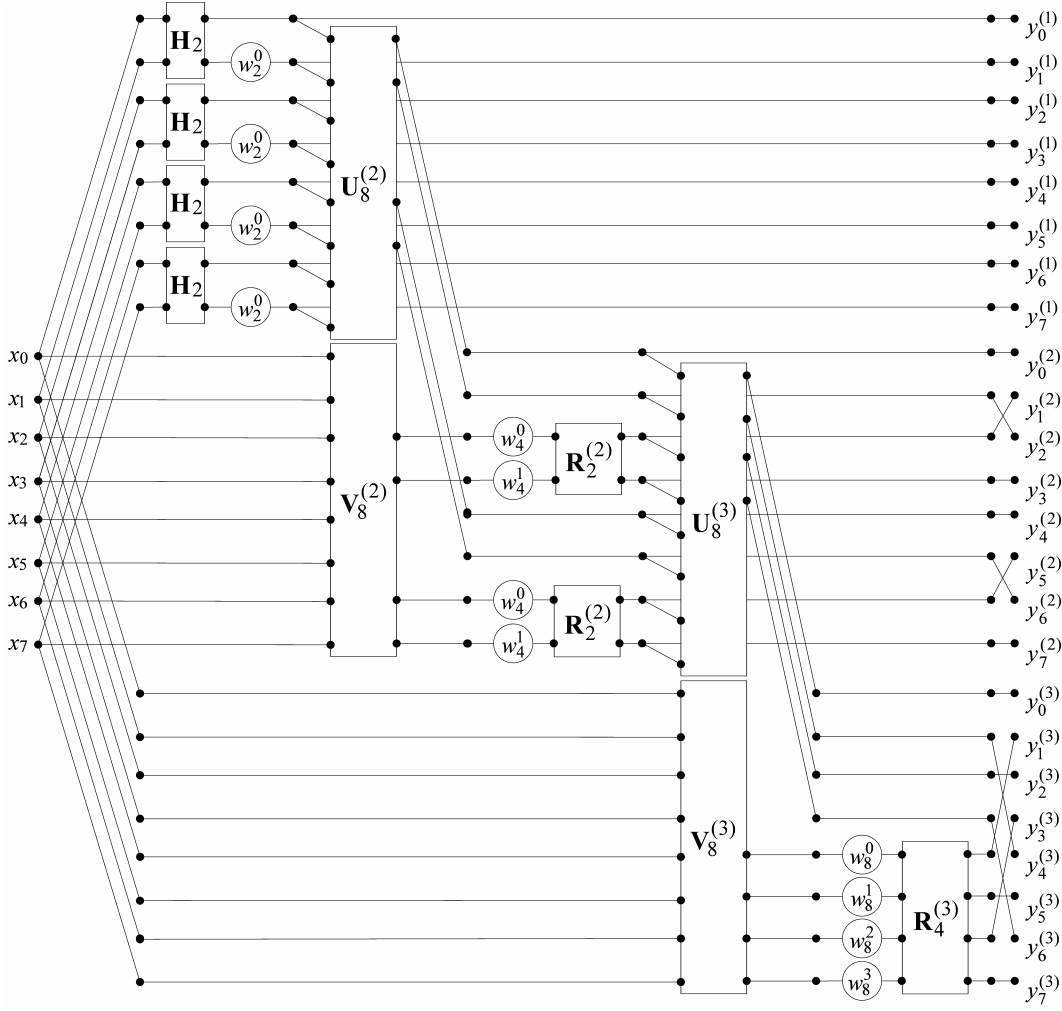

Fig. 2 – The data-flow diagram of the MrDFT calculation process for *N*=8 in accordance with the procedure given by the formula (5).

## 5. ANALYSIS OF COMPUTATIONAL COMPLEXITY

Now we consider the computational complexity of the proposed algorithm. The number of complex multiplications performed in the $i$-th iteration of the rationalized algorithm may be computed as follows:

$$\beta_\times^{(i)} = 2^{m-i} \cdot (2^{i-1} + 2^{i-2} \cdot \log_2 2^{i-1}) = (i+1) \cdot 2^{m-2}. \tag{6}$$

It should be emphasized that the formula (6) includes cases in relation to which one of the operands is "a trivial twiddle factor", which can be written as $w_n^0 = 1$ or $w_n^{n/4} = -\mathrm{j}$. For these values the complex multiplication can be reduced to no more than two change-sign operations of real and imaginary parts. Accordingly, it is possible to determine the number of complex multiplications performed in the $i$-th iteration with an exception of these "trivial calculations":

$$\widetilde{\beta}_\times^{(i)} = \begin{cases} 0 & \text{for } i = 1, \\ (i-2) \cdot 2^{m-2} & \text{for } i = 2, 3, \ldots, m. \end{cases} \tag{7}$$

Taking into account the formulas (6) and (7) it is possible to calculate the total number of complex multiplications performed in the algorithm. This can be done by summing $\beta_\times^{(i)}$ and $\widetilde{\beta}_\times^{(i)}$ over all iterations:

$$\beta_\times = \sum_{i=1}^{m} \beta_\times^{(i)} = m(m+3) \cdot 2^{m-3}, \quad \widetilde{\beta}_\times = \sum_{i=1}^{m} \widetilde{\beta}_\times^{(i)} = (m-1)(m-2) \cdot 2^{m-3}.$$

Taking into consideration the fact that the number of additions performed in each iteration is twice as big as the number of multiplications performed in the same iteration, it can be stated that the total number of additions equals $2 \cdot \beta_\times$.

## 6. CONCLUDING REMARKS

This article presents the efficient algorithm for the realization of the multiresolution discrete Fourier transform. The advantage of this algorithm is its ability to decrease the number of complex multiplications and additions. This advantage provides savings in hardware resources. The presented algorithm makes use of the results obtained at the previous level of spectral decomposition in the calculations which are performed in the current iteration. This leads to a reduction in the total amount of computation. As a result, the presented algorithm is characterized by a relevant reduction in the number of complex multiplications and additions when compared to the direct method, as well it provides almost 50% savings of arithmetical calculations when compared to the method based on the realizations of the traditional FFT algorithm at each resolution level. On the other hand, the presented algorithm is targeted at computing vectorization that will further reduce the delay of execution of the computational process. Due to the modular nature of the space-time structure of the computing process and the repeatability of its structural components, the algorithm can be efficiently implemented in various hardware or software platforms. Another advantage of the presented algorithm is a possibility of parallelization, so it is possible to achieve even greater reduction in the execution time. Given the fact that a multiresolution representation of the one-dimensional and two-dimensional signal is used in many practical DSP applications, we strongly repose hopes that the developed algorithm will be useful and effective technique of data analysis.